\documentclass[aps,reprint,twocolumn,superscriptaddress,amsmath,amssymb]{revtex4-2}

\usepackage{graphicx}
\usepackage{microtype}
\usepackage{siunitx}
\usepackage{booktabs}
\usepackage{xcolor}

\def\cucn{$\kappa$-(BEDT\--TTF)$_2$\-Cu$_2$(CN)$_3$}
\def\etcl{$\kappa$-(BEDT\--TTF)$_2$\-Cu[N(CN)$_2$]Cl}
\def\etbr{$\kappa$-(BEDT\--TTF)$_2$\-Cu[N(CN)$_2$]Br}
\def\eti{$\kappa$-(BEDT\--TTF)$_2$\-Cu[N(CN)$_2$]I}

\begin{document}
\title{Interacting electron spins in $\kappa$-(BEDT-TTF)$_2$Cu[N(CN)$_2$]I \\
investigated by ESR spectroscopy}
\author{Lena Nadine Majer}
\affiliation{1.~Physikalisches Institut, Universit\"{a}t Stuttgart, Pfaffenwaldring 57, 70569 Stuttgart, Germany}
\author{Bj\"{o}rn Miksch}
\affiliation{1.~Physikalisches Institut, Universit\"{a}t Stuttgart, Pfaffenwaldring 57, 70569 Stuttgart, Germany}
\author{Olga Iakutkina}
\affiliation{1.~Physikalisches Institut, Universit\"{a}t Stuttgart, Pfaffenwaldring 57, 70569 Stuttgart, Germany}
\author{Takuya Kobayashi}
\affiliation{Graduate School of Science and Engineering, Saitama University, Saitama, 338-8570, Japan}
\author{Atsushi Kawamoto}
\affiliation{Department of Condensed Matter Physics, Graduate School of Science, Hokkaido University, Sapporo 060-0810, Japan}
\author{Martin Dressel}
\affiliation{1.~Physikalisches Institut, Universit\"{a}t Stuttgart, Pfaffenwaldring 57, 70569 Stuttgart, Germany}

\date{\today}
\begin{abstract}
We performed angular and temperature-dependent electron-spin-resonance measurements in the quasi-two-dimensional organic conductor $\kappa$-(BEDT-TTF)$_2$Cu[N(CN)$_2$]I. The interlayer spin-diffusion is much weaker compared to the Cl- and Br-analogues,
which are antiferromagnetic insulator and paramagnetic metal, respectively;
$\kappa$-(BEDT-TTF)$_2$Cu[N(CN)$_2$]I behaves insulating when cooled below $T=\SI{200}{K}$.
A spin gap ($\Delta\approx\SI{18}{K}$) opens at low temperatures leading to a spin-singlet state.
Due to intrinsic disorder a substantial number of spins ($\sim\SI{1}{\percent}$) remains unpaired.
We observe additional signals below $T=\SI{4}{K}$ with a pronounced anisotropy indicating the presence of local magnetic moments coupled to some fraction of those unpaired spins.
\end{abstract}

\maketitle
\section{Introduction}
Among the quasi two-dimensional organic (BEDT-TTF)$_2X$ salts --
where BEDT-TTF denotes the organic cation bis(ethylenedithio)tetrathiafulvalene --
the $\kappa$-family draws particular attention, as slight modifications of the monovalent anions $X$
lead to a large variety of interesting ground states, reaching from Mott insulator to superconductor,
from antiferromagnet to quantum spin liquid \cite{Powell11,2020Dressel}.
One of the most prominent examples is the antiferromagnetic (afm) Mott insulator \etcl, which can be tuned to a $T_c=12.8$~K superconductor by applying  pressure of only 30~MPa \cite{91WANG-SynMetals}. When subsituting Cl with Br in the anion layer, the obtained \etbr\ is a weakly correlated metal and even a superconductor already at ambient pressure \cite{Kushch93,Faltermeier07,Dumm09,Yasin11}. 

This tendency, however, does not continue by going towards the isostructural \eti\ where an even larger halogen atom is incorporated in the polymeric anions;
the compound remains insulating at ambient conditions \cite{91WANG-SynMetals,91GEISER-PhysC}
and becomes superconducting with $T_c\approx 8$~K only when hydrostatic pressure of around 100~MPa is applied \cite{00Tanatar-PhysRevB,Tanatar02}.
From early on it was suggested that the metallic state is suppressed by structural defects \cite{91GEISER-PhysC} rather than electronic correlations \cite{Kanoda97a,Mori99b}.
Interestingly, when disorder is intentionally introduced to the afm insulator \etcl\ by massive x-ray irradiation, the magnetic order is suppressed towards a state resembling the quantum spin liquid state in \cucn\ \cite{15Furukawa-PhysRevLett}.

Here we tackle the question about the magnetic ground state in \eti. Triggered by our recent broadband investigations of the quantum spin liquid Mott insulator \cucn\ utilizing electron spin resonance \cite{20Miksch}, we perform comprehensive temperature and angular-dependent ESR measurements on \eti\ single crystals. Due to the strong effect of disorder, our findings differ rather strongly from the ones previously obtained on
\etcl\ and \etbr\ \cite{09Antal-PhysRevLett,10Antal-PhysicaB,11Antal-PhysRevB}.

\section{Experimental details}
\begin{figure}[b]
\centering
	\includegraphics[width=0.5\columnwidth]{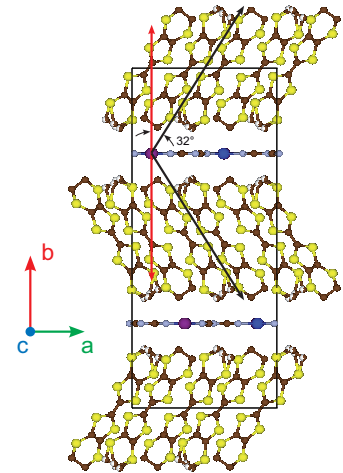}
    \caption{Structure of \eti\ viewed along the $c$-direction.
    In subsequent layers the long axis of the BEDT-TTF dimers is tilted by an alternating angle of $\pm 32^\circ$ with respect to the $b$-direction.
    Here the yellow, brown, blue, purple and grey color denotes sulfur, carbon, copper, iodine and nitrogen atoms, respectively. (The atomic positions were taken from \cite{91GEISER-PhysC})}
  \label{fig:Structure}
\end{figure}
Single crystals of the low-dimensional organic conductor \eti\ were prepared electrochemically as described in Ref.~\onlinecite{19Kobayashi-PhysRevB}.
$\kappa$-(BEDT-TTF)$_2$Cu[N(CN)$_2$]$X$ with $X$ = Cl, Br and I are isostructural and crystallize in the space group Pnma \cite{91GEISER-PhysC}.
As illustrated in Fig.~\ref{fig:Structure}, the BEDT-TTF dimers are tilted with respect to the stacking direction with alternating incliniation.
Two adjacent organic layers are related by mirror symmetry; in each layer there are four inequivalent BEDT-TTF molecules per unit cell.
The BEDT-TTF layers are stacked along the $b$-direction separated by polymeric Cu[N(CN)$_2$]I anions.
The crystal structure was confirmed by comparing the lattice parameters measured by x-ray on a sample from the same batch with published data \cite{91GEISER-PhysC}.
The crystals are typically flat platelets of mm size with rhombic shape normal to the $b$-axis.
The orientation was done via infrared reflection measurements \cite{Dressel04}.
The temperature-dependent dc resistivity was measured along the crystallographic $b$-axis to characterize the samples using the standard four-probe method.

The electron spin resonance (ESR) measurements were performed on a continuous-wave spectrometer (Bruker EMXplus) in X-band (approximately \SI{9.5}{GHz}).
The system is equipped with a goniometer enabling angular dependent measurements.
Experiments from room temperature down to $T=\SI{4}{K}$ were possible using a continuous He-gas flow cryostat (Oxford instruments ESR 900). In order to reach even lower temperatures $T>\SI{2}{K}$ a pumped cryostat (ESR 910) is employed.
We observed the principal axes of the $g$-tensor to correspond to the $a$, $b$ and $c$-directions of the lattice (Fig.~\ref{fig:Structure}).

\section{Results and Discussion}
\subsection{Transport and Magnetic Characterization}
\label{sec:characterization}
\begin{figure}
\centering
	\includegraphics[width=1\columnwidth]{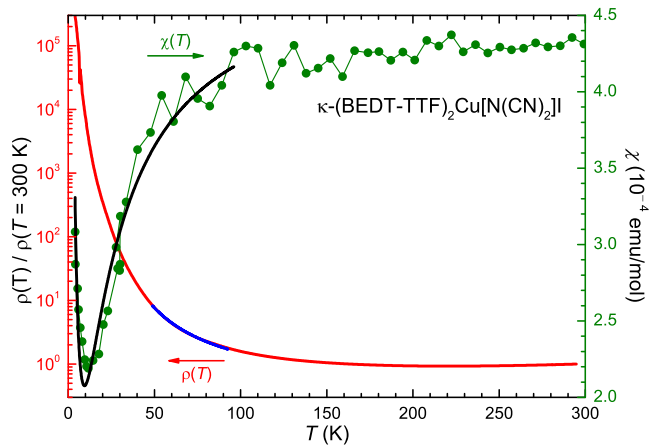}
    \caption{Temperature-dependent resistivity $\rho(T)$ measured along the $b$-axis of \eti\ and normalized to the room temperature value (red line, left axis). The blue line indicates the activated behavior in a restricted temperature range. The green dots correspond to the spin susceptibility $\chi(T)$ as extracted from the ESR spectra with the magnetic field along $b$; the data are normalized to the molar susceptibility \cite{19Kobayashi-PhysRevB} at \SI{200}{K} (right axis).
    The low temperature spin susceptibility can be modelled by an activated behavior due to the formation of a singlet state accompanied by the opening of a spin gap in cobination with a Curie contribution due to the remaining unpaired spins (black line).}
  \label{fig:DC-SQUID-ESR-Mix}
\end{figure}

A weakly metallic behavior is present in \eti\ only above
the shallow minimum around $T=\SI{200}{K}$ in the dc resistivity (Fig.~\ref{fig:DC-SQUID-ESR-Mix}).
Upon further cooling the sample develops an insulating behavior.
The anisotropy of the resistivity is temperature-independent.
It is not possible to describe the complete behavior of $\rho(T)$ with a single activation energy $\Delta$; in the range between 50 and \SI{90}{K} an Arrhenius fit (blue line) yields  approximately $\Delta \approx \SI{160}{K}$ with a slight decrease towards lower temperatures  in accord with previous reports \cite{91WANG-SynMetals,00Tanatar-PhysRevB}.
As this increase continues towards the lowest temperatures along the $b$-axis,
no traces of superconductivity are detected down to $T=\SI{2}{K}$, confirming the results of other groups \cite{91GEISER-PhysC,00Tanatar-PhysRevB,Tanatar02}.

The spin susceptibility of \eti\ is obtained from fits to the ESR spectra; the results are displayed in Fig.~\ref{fig:DC-SQUID-ESR-Mix} as a function of temperature.
The data were normalized to the molar susceptibility extracted from SQUID measurements \cite{19Kobayashi-PhysRevB}.

In the high-temperature range the spin susceptibility is nearly temperature independent.
Coincident with the strong increase of the resistivity $\rho(T)$ below \SI{90}{K}, the spin susceptibility $\chi_S(T)$ starts to decrease,
in accord to \cite{92KATAEV-PhysicaB,92KATAEV-SolStatCom,19Kobayashi-PhysRevB}.
Eventually the spin susceptibility exhibits a minimum at low-temperature.
The Curie like increase at low $T$ indicates the presence of localized unpaired spins.
Our findings are consistent with magnetization data 
  that exhibit a similar increase upon cooling below \SI{7}{K} \cite{19Kobayashi-PhysRevB}.

The low-temperature behavior can be modeled taking into account two contributions
\begin{equation}
	\chi (T) = A\cdot \exp\left\{-\Delta / T\right\} + C/T \quad ;
\end{equation}
the activated behavior due to the pairing of spins into singlets
resulting in a spin gap $\Delta$; and a Curie contribution accounting for the unpaired spins. Here, $C$ denotes the Curie constant.
A fit to the data shown in Fig.~\ref{fig:DC-SQUID-ESR-Mix} results in the parameters $A=4.9 \cdot 10^{-4}$ ~emu/mol, $\Delta=\SI{18.2}{K}$ and $C=\SI{0.0013}{emu K/mol}$.
Thus, about 1 $\%$ of the total spins remain unpaired at low temperatures contributing to the Curie tail.

\subsection{Effects of Disorder}
The Hubbard model commonly used to describe the electronic properties of the BEDT-TTF salts, generally predicts some antiferromagnetic order at low temperature in the Mott-insulating phase. Disorder can however prevent long-range magnetic order. On short length scales the formation of spin $S=0$ valence bonds can however still be favored due to the antiferromagnetic exchange.
Previous comprehensive $^{13}$C-NMR studies on \eti\  indeed indicate the presence of antiferromagnetic fluctuations at low temperatures despite the lack of long-range magnetic order \cite{19Kobayashi-PhysRevB}. Similar observations have been made for x-ray irradiated samples of \etcl\ \cite{15Furukawa-PhysRevLett}.
In this case the irradiation creates disorder within the sample suppressing the long-range magnetic order.
Many of the properties received this way are similar to those of spin liquid compounds \cite{15Furukawa-PhysRevLett}. As the main source of disorder the conformation of ethylene endgroups has been proposed. The ethylene groups of the BEDT-TTF donor molecules in the sister compounds \etcl\ and \etbr\ are ordered as the temperature is reduced. In contrast, one of the ethylene groups of BEDT-TTF in \eti\ is disordered \cite{91GEISER-PhysC}.

Most of the spins dimerize into valence bonds, leading to an effective spin singlet state, as indicated by the observed spin gap, at low temperature. Disorder can cause random pinning of the valence bonds resulting in localized unpaired spins hosted in between the dimers. These unpaired defect spins are responsible for the low-temperature signal observed in all measured crystals stemming from multiple batches and laboratoriesf. In the results shown, the defect spins account to approximately $1\,\%$ of the spins observed above the pairing temperature.

\subsection{g-Factor Anisotropy}

\begin{figure}
\centering
	\includegraphics[width=1\columnwidth]{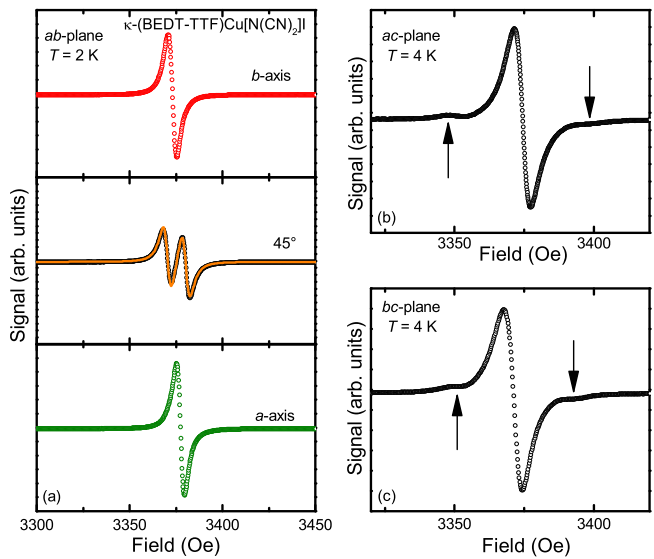}
    \caption{(a)~ESR spectra recorded at $T=\SI{2}{K}$ with the magnetic field in the $ab$-plane. For orientations midway between the $a$- and $b$-axes, a splitting of the signal is visible (shown here for \SI{45}{\degree}). The spectra in the right column demonstrates that also with the external magnetic field in the $ab$- and $bc$-planes additional signals can be observed (black arrows):~(b)~$ab$-plane when rotated \SI{20}{\degree} with respect to the $a$-axis at $T=\SI{4}{K}$; (c)~$bc$-plane at \SI{16}{\degree} with respect to the $b$-axis.}
 \label{fig:Signals}
\end{figure}
The $g$-values of \eti\ obtained from our measurements show a slight anisotropy: $g_a = 2.006$, $g_b = 2.009$ and $g_c = 2.006$;
no significant change is observed from ambient temperature down to $T=\SI{4}{K}$.
A comparable out-of plane anisotropy of the $g$-factor was reported by Kataev {\it et al.} in an extensive ESR study of \etbr\ and \eti\ \cite{92KATAEV-PhysicaB,92KATAEV-SolStatCom}.

More important, however, are the observations extracted from our comprehensive angular-dependent ESR investigations
on large \eti\ single crystals, which yield some rather unexpected results.
Fig.~\ref{fig:Signals}(a) displays ESR spectra measured  at $T=\SI{2}{K}$ with the magnetic field in different orientations within the $ab$-plane.
Parallel to the $b$-axis (red line) and the $a$-axis (green line), we observed a single absorption line at slightly different resonance fields.
The signal, however, does not shift continuously upon rotating the sample, as expected for a simple $g$-tensor.
We instead observed a pronounced splitting of the spectrum for the diagonal \SI{45}{\degree} direction (black curve).
This splitting was only found in the $ab$-plane and can be explained by the alternating orientation of the BEDT-TTF molecules in adjacent layers (Fig. \ref{fig:Structure}).

If a crystal contains chemically equivalent but crystallographically different molecules in adjacent layers ($\alpha$ and $\beta$), as is the case for \eti,
the ESR spectrum is composed of two lines with $g$-factors corresponding to $g_\alpha$ and $g_\beta$-tensors;
the respective features occur at different resonance frequencies $\nu_\alpha$ and $\nu_\beta$.
The separation $(\nu_\alpha-\nu_\beta)$ is proportional to the external magnetic field and described by
\begin{equation}
	\label{eq:diff_in_Frequ}
    \nu_\alpha-\nu_\beta=(g_\alpha-g_\beta)\mu_\text{B}B/h \quad ;
\end{equation}
here $\mu_\text{B}$ is the Bohr magneton ($\mu_B=\frac{e \hbar}{2 m_e}$) and $h = 2 \pi \hbar$ the Planck constant.
Along the high symmetry directions $a$ and $b$, the two lines collapse into one.

The full $ab$-anisotropy is plotted in Fig.~\ref{fig:rotation}(a).
Both lines show an identical anisotropy, but are shifted in phase by approximately \SI{65}{\degree}.
The minimum resonance field of the two lines occurs symmetrically at \SI{\pm 32}{\degree} off the $b$-axis.
This closely resembles the crystal structure of the compound, where in adjacent layers the BEDT-TTF molecules are tilted by an angle of \SI{32}{\degree} in alternating directions within the $ab$-plane as depicted in Fig.~\ref{fig:Structure}.
Symmetry requires that both lines coincide along the $a$- and $b$-axes.
Very similar observations have been reported for the sister compounds \etcl\ and \etbr\ \cite{09Antal-PhysRevLett,11Antal-PhysRevB}.

\subsection{Spin Diffusion}
As the individual BEDT-TTF dimers are not completely isolated, the spins are subject to exchange interaction.
The simple picture of two ESR lines stemming from the crystallographically inequivalent layers $\alpha$ and $\beta$ does not hold anymore if the interlayer spin hopping $\nu_\perp$ is comparable to the separation of the lines $|\nu_\alpha -\nu_\beta |$ \footnote{This follows the elegant phenomenological description of lineshapes by Kubo \cite{Kubo69}; the two-state jump model it is also applied to model the splitting of vibrational lines in charge-order compounds with significant charge fluctuations \cite{Girlando12}}.
Three cases can be distinguished \cite{Thesis-Anges-Dissertation}: If $\nu_\perp \ll |\nu_\alpha -\nu_\beta|$, the coupling is weak; hence two slightly modified lines can be resolved. For $\nu_\perp \approx |\nu_\alpha -\nu_\beta |$ the distortion becomes more pronounced. And in the limit $\nu_\perp \gg |\nu_\alpha -\nu_\beta |$,  one exchange-narrowed line remains at the central frequency $(\nu_\alpha + \nu_\beta )/2$.

In the central panel of Fig.~\ref{fig:Signals}(a), we show a fit of two uncoupled Lorentzian signals (orange line) to the split signal in the diagonal direction.
The model does not yield a substantial deviation from the measured signal, indicating only a weak coupling between the two lines.

\begin{figure}
\centering
	\includegraphics[width=1\columnwidth]{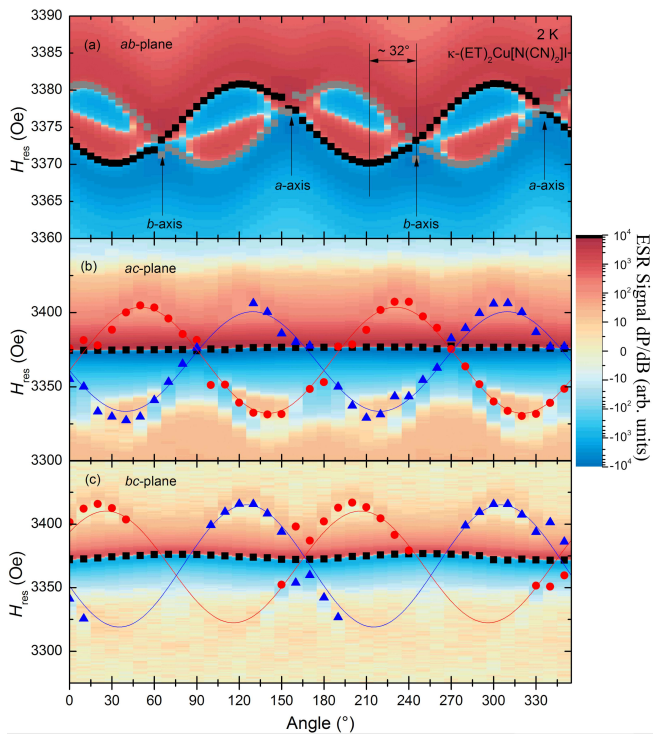}
    \caption{(a) The anisotropy of the ESR signal of \eti\ measured in the $ab$-plane at a temperature of $T=\SI{2}{K}$. The positions of the resonance fields are represented by the grey and black squares.
    (b) When rotating the crystal within the $ac$-plane the anisotropy of the main signal (indicated by the black squares)
     is meager compared to the angular dependence observed in the position of the small dips, shown by red dots and blue triangles.
     (c) A similar observation is made when rotating in the $bc$-plane. Besides the small angular dependence of the main line, we find a strong anisotropy of the two dips. Note the extended scales for panels (b) and (c) compared to panel (a). The background color corresponds to the ESR signal intensity ${\rm d}P/{\rm d}B$.}
  \label{fig:rotation}
\end{figure}

Although there are not sufficient data points to extract even a qualitative relation, we propose a sort of thermally activated hopping of the spins among the organic layers.
When carefully inspecting the spectra from Fig.~\ref{fig:rotation}(a), we can resolve two lines when they differ by about \SI{1.5}{Oe} in resonance field.
This allows us to estimate an upper limit for the interlayer hopping time for the spins $\frac{1}{2 \pi \nu_{\perp}}=\tau_{\perp} \geq 3.7 \times 10^{-8}$~s by following the procedure suggested by Antal {\it et al.} \cite{11Antal-PhysRevB,09Antal-PhysRevLett}.
For $T=\SI{4}{K}$ only lines separated by more than \SI{2}{Oe} can be distinguished, hence we obtain $\tau_{\perp} \geq 2.9 \times 10^{-8}$~s.
For higher temperatures, it becomes impossible to identify the features reliably since the signals broaden significantly.
At a temperature of \SI{2}{K} the splitting can be better resolved than at $T=\SI{4}{K}$. This can be explained by the larger linewidth at \SI{4}{K} either due to the lifetime broadening or due to a thermally activated hopping of the spin carrier.

In the case of \etcl\ and \etbr, Antal {\it et al.} could not observe any splitting by X-band spectroscopy;
however, employing high-frequency ESR at \SI{111.2}{GHz} and \SI{222.4}{GHz} a clear splitting was identified already at $T=\SI{250}{K}$ with no appreciable temperature dependence \cite{09Antal-PhysRevLett,11Antal-PhysRevB}.
The authors estimated an upper limit for the interlayer spin hopping time with $\tau_{\perp} \geq 1.4 \times 10^{-9}$~s --~one order of magnitude smaller than our findings in \eti.
Applying hydrostatic pressure, and thus shrinking the unit cell of \etbr\ and \etcl\ at \SI{250}{K} leads to an increase in the inter layer hopping rate \cite{Thesis-Anges-Dissertation,10Antal-PhysicaB}.
If the pressure is sufficiently high, the two ESR lines merge into one, even at higher frequencies.
The length of the unit cell of the \eti\ sample in the interlayer stacking direction $b$ has a larger value (30.356 \AA\ at $T$ = 295 K) than in the case of the \etcl\ and \etbr\ samples \cite{91GEISER-PhysC}, the layers can therefore be expected to be more decoupled which is also evidenced by our ESR results.

The electronic anisotropy can be estimated (in analogy to the sister components \etbr\ and \etcl\ in Ref. \cite{09Antal-PhysRevLett}) via 
\begin{equation}
\frac{\sigma_\Vert}{\sigma_\perp}=\left(\frac{t_\Vert}{t_\perp}\right)^2 \cdot \left(\frac{\ell}{b}\right)^2
\quad;
\end{equation}
here $\ell$ corresponds to the mean free path and $b$ to the distance between two layers. Using $t_{\perp}=\frac{\hbar}{\sqrt{2 \tau_{\Vert}\tau_{\perp}}}$ we can estimate a value  $t_{\perp}$ for the coupling between the layers \cite{92Kumar-PhysRevB,09Antal-PhysRevLett}.
Assuming a value of $t_\Vert = 0.1$~\text{eV} for the intralayer transfer integral between the BEDT-TTF dimers and using $\nu_\text{F} \approx 10^5 \text{m/s}$ for the Fermi velocity, the anisotropy for \etbr\  and \etcl\ was estimated to $10^5-10^6$ \cite{09Antal-PhysRevLett}. Applying the same estimate to \eti, we obtain an anisotropy of ${\sigma_\Vert}/{\sigma_\perp} = 10^6$.
In other words, the anisotropy determined from our ESR measurements has the same order of magnitude as for the Cl- and Br sibling  compounds.   If this value is compared to the typical dc conductivity anisotropy of this sample family, the value calculated here is significantly larger. For \etcl\ and \etbr\ the dc conductivity anisotropy is typically in the range of $10^2$ to $10^3$ \cite{10Antal-PhysicaB,06Zverev,92Buraviv}.

\subsection{Low-temperature anisotropy}
By now we confined ourselves to ESR experiments along the $ab$-plane of \eti\
where the small anisotropy is determined by the $g$-factor of the tilted the BEDT-TTF molecules, as indicated in Fig.~\ref{fig:Structure}.
Fig.~\ref{fig:Signals}(b) displays an exemplary ESR spectrum measured in the $ac$-plane at $T=\SI{4}{K}$ recorded when rotating the sample by \SI{20}{\degree} with respect to the $a$-axis. Besides the main signal, two additional features are observed on both sides that shift in resonance field upon rotation.
A similar behavior is found in the $bc$-plane as presented Fig.~\ref{fig:Signals}(c) for an orientation of \SI{16}{\degree} with respect to the $b$-axis.
In certain directions only one feature can be identified due to distortion in the vicinity of the main signal.
In general, the additional lines were more pronounced in the $ac$-plane compared to the $bc$-plane;
within the $ab$-plane we could not observe such features.

The full anisotropy at $T=\SI{2}{K}$ is shown Fig.~\ref{fig:rotation}(b,c) in the $ac$- and  $bc$-planes of \eti\ single crystals.
The resonance field of the main signal is indicated by black squares. The red dots and the blue triangles represent the resonance field of the two additional features.
The anisotropy of these lines is approximately two orders of magnitude larger than the anisotropy of the main signal given by the slightly anisotropic $g$-tensor due to spin-orbit coupling. The difference of the maximum and minimum resonance field of the main signal is only  $\Delta H_\text{res, main}\approx 4$ ~Oe, whereas for the
additional lines a much larger anisotropy $\Delta H_\text{res, imperfections}$ of about 80 to 100 ~Oe is observed.

As discussed in section \ref{sec:characterization}, a spin singlet state is formed at low temperatures as evidenced by the opening of a spin gap; about \SI{1}{\percent} of the spins remain unpaired, contributing to the observed Curie tail.
However, the number of spins associated with the additional anisotropic ESR feature observed at low temperatures is substantially smaller; for the crystals presented in Figs.~\ref{fig:Signals} and \ref{fig:rotation} it amounts to only \SI{270}{ppm}. This implies that only \SI{2.7}{\percent} of the unpaired spins contribute to the additional signal.

We suggest that the additional signal is related to magnetic imperfections:
some of the unpaired defect spins are formed in the vicinity of these imperfections.
Coupling to the imperfections, e.g.\ via dipolar interaction, can lead to the internal field necessary for the observed large anisotropy.
As it is known that the synthesis of \eti\ is quite demanding compared to the Cl and Br-analogues, magnetic imperfections are likely present in the crystals \cite{Kushch01}.

This result is in accord with recent ESR investigations on the quantum spin liquid candidate \cucn\ \cite{20Miksch}. At low-temperatures Miksch {\it et al.} identified a second contribution that splits upon cooling and reveals a rather pronounced anisotropy. From the sample-, temperature-, field-, and angular-dependences, this feature was linked to localized magnetic moments, which do not participate in the formation of spin singlets. These orphan spins couple to Cu$^{2+}$ impurities in the anion layer via dipole-dipole interaction.
\\

\section{Conclusion}
We can identify a splitting of the ESR signal caused by the different layers, already at X-band frequencies, indicating an even smaller interlayer spin hopping rate compared to \etcl\ and \etbr. In summary, our ESR investigations on  \eti\ single crystals yield an upper limit of the interlayer spin exchange, given by $\tau_{\perp} \geq 3.7 \times 10^{-8}$~s at $T=\SI{2}{K}$.

At low temperatures a spin-singlet state with a gap $\Delta\approx\SI{18}{K}$ is formed.
In addition, our ESR spectra indicate that a substantial amount of the spins ($\approx$ 1 $\%$) remains unpaired, contributing significantly to the low-temperature susceptiblility.
In combination with magnetic imperfections a small number of these unpaired spins is responsible for the observed additional strongly anisotropic ESR signal.

\acknowledgments
This study was initiated by Y. Saito bringing the crystals to Stuttgart.
For comparison, crystals were  provided by J.A. Schlueter and A.A. Bardin.
We are very grateful to  A. Pustogow for fruitful discussions, and also thank
{\'A}. Antal, B. N{\'a}fr{\'a}di, L. Forr{\'o} and A. J{\'a}nossy.
The work was supported by the Deutsche Forschungsgemeinschaft (DFG) via DR228/39-3 and DR228/52-1 and partially by the Japan Society for the Promotion of Science (JSPS) KAKENHI Grant Numbers 19K03758 and 20K14401.

\bibliography{references}

\begin{thebibliography}{29}%
\makeatletter
\providecommand \@ifxundefined [1]{%
 \@ifx{#1\undefined}
}%
\providecommand \@ifnum [1]{%
 \ifnum #1\expandafter \@firstoftwo
 \else \expandafter \@secondoftwo
 \fi
}%
\providecommand \@ifx [1]{%
 \ifx #1\expandafter \@firstoftwo
 \else \expandafter \@secondoftwo
 \fi
}%
\providecommand \natexlab [1]{#1}%
\providecommand \enquote  [1]{``#1''}%
\providecommand \bibnamefont  [1]{#1}%
\providecommand \bibfnamefont [1]{#1}%
\providecommand \citenamefont [1]{#1}%
\providecommand \href@noop [0]{\@secondoftwo}%
\providecommand \href [0]{\begingroup \@sanitize@url \@href}%
\providecommand \@href[1]{\@@startlink{#1}\@@href}%
\providecommand \@@href[1]{\endgroup#1\@@endlink}%
\providecommand \@sanitize@url [0]{\catcode `\\12\catcode `\$12\catcode
  `\&12\catcode `\#12\catcode `\^12\catcode `\_12\catcode `\%12\relax}%
\providecommand \@@startlink[1]{}%
\providecommand \@@endlink[0]{}%
\providecommand \url  [0]{\begingroup\@sanitize@url \@url }%
\providecommand \@url [1]{\endgroup\@href {#1}{\urlprefix }}%
\providecommand \urlprefix  [0]{URL }%
\providecommand \Eprint [0]{\href }%
\providecommand \doibase [0]{https://doi.org/}%
\providecommand \selectlanguage [0]{\@gobble}%
\providecommand \bibinfo  [0]{\@secondoftwo}%
\providecommand \bibfield  [0]{\@secondoftwo}%
\providecommand \translation [1]{[#1]}%
\providecommand \BibitemOpen [0]{}%
\providecommand \bibitemStop [0]{}%
\providecommand \bibitemNoStop [0]{.\EOS\space}%
\providecommand \EOS [0]{\spacefactor3000\relax}%
\providecommand \BibitemShut  [1]{\csname bibitem#1\endcsname}%
\let\auto@bib@innerbib\@empty
\bibitem [{\citenamefont {Powell}\ and\ \citenamefont
  {McKenzie}(2011)}]{Powell11}%
  \BibitemOpen
  \bibfield  {author} {\bibinfo {author} {\bibfnamefont {B.~J.}\ \bibnamefont
  {Powell}}\ and\ \bibinfo {author} {\bibfnamefont {R.~H.}\ \bibnamefont
  {McKenzie}},\ }\bibfield  {title} {\bibinfo {title} {Quantum frustration in
  organic mott insulators: From spin liquids to unconventional
  superconductors},\ }\href@noop {} {\bibfield  {journal} {\bibinfo  {journal}
  {Rep. Progr. Phys.}\ }\textbf {\bibinfo {volume} {74}},\ \bibinfo {eid}
  {056501} (\bibinfo {year} {2011})}\BibitemShut {NoStop}%
\bibitem [{\citenamefont {Dressel}\ and\ \citenamefont
  {Tomić}(2020)}]{2020Dressel}%
  \BibitemOpen
  \bibfield  {author} {\bibinfo {author} {\bibfnamefont {M.}~\bibnamefont
  {Dressel}}\ and\ \bibinfo {author} {\bibfnamefont {S.}~\bibnamefont
  {Tomić}},\ }\bibfield  {title} {\bibinfo {title} {{Molecular quantum
  materials: electronic phases and charge dynamics in two-dimensional organic
  solids}},\ }\href {https://doi.org/10.1080/00018732.2020.1837833} {\bibfield
  {journal} {\bibinfo  {journal} {Advances in Physics}\ }\textbf {\bibinfo
  {volume} {69}},\ \bibinfo {pages} {1} (\bibinfo {year} {2020})}\BibitemShut
  {NoStop}%
\bibitem [{\citenamefont {Wang}\ \emph {et~al.}(1991)\citenamefont {Wang},
  \citenamefont {Carlson}, \citenamefont {Geiser}, \citenamefont {Kini},
  \citenamefont {Schultz}, \citenamefont {Williams}, \citenamefont
  {Montgomery}, \citenamefont {Kwok}, \citenamefont {Welp}, \citenamefont
  {Vandervoort}, \citenamefont {Boryschuk}, \citenamefont {Crouch},
  \citenamefont {Kommers}, \citenamefont {Watkins}, \citenamefont {Schriber},
  \citenamefont {Overmyer}, \citenamefont {Jung}, \citenamefont {Novoa},\ and\
  \citenamefont {Whangbo}}]{91WANG-SynMetals}%
  \BibitemOpen
  \bibfield  {author} {\bibinfo {author} {\bibfnamefont {H.}~\bibnamefont
  {Wang}}, \bibinfo {author} {\bibfnamefont {K.}~\bibnamefont {Carlson}},
  \bibinfo {author} {\bibfnamefont {U.}~\bibnamefont {Geiser}}, \bibinfo
  {author} {\bibfnamefont {A.}~\bibnamefont {Kini}}, \bibinfo {author}
  {\bibfnamefont {A.}~\bibnamefont {Schultz}}, \bibinfo {author} {\bibfnamefont
  {J.}~\bibnamefont {Williams}}, \bibinfo {author} {\bibfnamefont
  {L.}~\bibnamefont {Montgomery}}, \bibinfo {author} {\bibfnamefont
  {W.}~\bibnamefont {Kwok}}, \bibinfo {author} {\bibfnamefont {U.}~\bibnamefont
  {Welp}}, \bibinfo {author} {\bibfnamefont {K.}~\bibnamefont {Vandervoort}},
  \bibinfo {author} {\bibfnamefont {S.}~\bibnamefont {Boryschuk}}, \bibinfo
  {author} {\bibfnamefont {A.}~\bibnamefont {Crouch}}, \bibinfo {author}
  {\bibfnamefont {J.}~\bibnamefont {Kommers}}, \bibinfo {author} {\bibfnamefont
  {D.}~\bibnamefont {Watkins}}, \bibinfo {author} {\bibfnamefont
  {J.}~\bibnamefont {Schriber}}, \bibinfo {author} {\bibfnamefont
  {D.}~\bibnamefont {Overmyer}}, \bibinfo {author} {\bibfnamefont
  {D.}~\bibnamefont {Jung}}, \bibinfo {author} {\bibfnamefont {J.}~\bibnamefont
  {Novoa}},\ and\ \bibinfo {author} {\bibfnamefont {M.-H.}\ \bibnamefont
  {Whangbo}},\ }\bibfield  {title} {\bibinfo {title} {{New $\kappa$-phase
  materials, $\kappa$-(ET)$_2$Cu[N(CN)$_2$]$X$. $X$=Cl, Br and I. The
  synthesis, structure and superconductivity above 11 K in the Cl ($T_c = 12.8$
  K, 0.3 kbar) and Br ($T_c = 11.6$ K) salts}},\ }\href
  {https://doi.org/https://doi.org/10.1016/0379-6779(91)91996-N} {\bibfield
  {journal} {\bibinfo  {journal} {Synth. Metals}\ }\textbf {\bibinfo {volume}
  {42}},\ \bibinfo {pages} {1983 } (\bibinfo {year} {1991})}\BibitemShut
  {NoStop}%
\bibitem [{\citenamefont {Kushch}\ \emph {et~al.}(1993)\citenamefont {Kushch},
  \citenamefont {Buravov}, \citenamefont {Khomenko}, \citenamefont {Yagubskii},
  \citenamefont {Rosenberg},\ and\ \citenamefont {Shibaeva}}]{Kushch93}%
  \BibitemOpen
  \bibfield  {author} {\bibinfo {author} {\bibfnamefont {N.}~\bibnamefont
  {Kushch}}, \bibinfo {author} {\bibfnamefont {L.}~\bibnamefont {Buravov}},
  \bibinfo {author} {\bibfnamefont {A.}~\bibnamefont {Khomenko}}, \bibinfo
  {author} {\bibfnamefont {E.}~\bibnamefont {Yagubskii}}, \bibinfo {author}
  {\bibfnamefont {L.}~\bibnamefont {Rosenberg}},\ and\ \bibinfo {author}
  {\bibfnamefont {R.}~\bibnamefont {Shibaeva}},\ }\bibfield  {title} {\bibinfo
  {title} {{{Novel organic superconductor
  $\kappa$-(ET)$_2$Cu[N(CN)$_2$]Cl$_{0.5}$Br$_{0.5}$ with $T_c \approx
  11.3$~K}}},\ }\href
  {https://doi.org/https://doi.org/10.1016/0379-6779(93)90887-3} {\bibfield
  {journal} {\bibinfo  {journal} {Synthetic Metals}\ }\textbf {\bibinfo
  {volume} {53}},\ \bibinfo {pages} {155 } (\bibinfo {year}
  {1993})}\BibitemShut {NoStop}%
\bibitem [{\citenamefont {Faltermeier}\ \emph {et~al.}(2007)\citenamefont
  {Faltermeier}, \citenamefont {Barz}, \citenamefont {Dumm}, \citenamefont
  {Dressel}, \citenamefont {Drichko}, \citenamefont {Petrov}, \citenamefont
  {Semkin}, \citenamefont {Vlasova}, \citenamefont {Me\'{z}i\`{e}re},\ and\
  \citenamefont {Batail}}]{Faltermeier07}%
  \BibitemOpen
  \bibfield  {author} {\bibinfo {author} {\bibfnamefont {D.}~\bibnamefont
  {Faltermeier}}, \bibinfo {author} {\bibfnamefont {J.}~\bibnamefont {Barz}},
  \bibinfo {author} {\bibfnamefont {M.}~\bibnamefont {Dumm}}, \bibinfo {author}
  {\bibfnamefont {M.}~\bibnamefont {Dressel}}, \bibinfo {author} {\bibfnamefont
  {N.}~\bibnamefont {Drichko}}, \bibinfo {author} {\bibfnamefont
  {B.}~\bibnamefont {Petrov}}, \bibinfo {author} {\bibfnamefont
  {V.}~\bibnamefont {Semkin}}, \bibinfo {author} {\bibfnamefont
  {R.}~\bibnamefont {Vlasova}}, \bibinfo {author} {\bibfnamefont
  {C.}~\bibnamefont {Me\'{z}i\`{e}re}},\ and\ \bibinfo {author} {\bibfnamefont
  {P.}~\bibnamefont {Batail}},\ }\bibfield  {title} {\bibinfo {title}
  {{Bandwidth-controlled Mott transition in {\rm
  $\kappa$-(BEDT-TTF)$_2$\-Cu[N(CN)$_2$]Br$_{x}$Cl$_{1-x}$}: Optical studies of
  localized charge excitations}},\ }\href@noop {} {\bibfield  {journal}
  {\bibinfo  {journal} {Phys. Rev. B}\ }\textbf {\bibinfo {volume} {76}},\
  \bibinfo {eid} {165113} (\bibinfo {year} {2007})}\BibitemShut {NoStop}%
\bibitem [{\citenamefont {Dumm}\ \emph {et~al.}(2009)\citenamefont {Dumm},
  \citenamefont {Falter\-meier}, \citenamefont {Drichko}, \citenamefont
  {Dressel}, \citenamefont {Me\'{z}i\`{e}re},\ and\ \citenamefont
  {Batail}}]{Dumm09}%
  \BibitemOpen
  \bibfield  {author} {\bibinfo {author} {\bibfnamefont {M.}~\bibnamefont
  {Dumm}}, \bibinfo {author} {\bibfnamefont {D.}~\bibnamefont {Falter\-meier}},
  \bibinfo {author} {\bibfnamefont {N.}~\bibnamefont {Drichko}}, \bibinfo
  {author} {\bibfnamefont {M.}~\bibnamefont {Dressel}}, \bibinfo {author}
  {\bibfnamefont {C.}~\bibnamefont {Me\'{z}i\`{e}re}},\ and\ \bibinfo {author}
  {\bibfnamefont {P.}~\bibnamefont {Batail}},\ }\bibfield  {title} {\bibinfo
  {title} {{Band\-width-controlled Mott transition in {\rm
  $\kappa$-(BEDT-TTF)$_2$\-Cu[N(CN)$_2$]Br$_x$Cl$_{1-x}$}: Optical studies of
  correlated carriers}},\ }\href@noop {} {\bibfield  {journal} {\bibinfo
  {journal} {Phys. Rev. B}\ }\textbf {\bibinfo {volume} {79}},\ \bibinfo {eid}
  {195106} (\bibinfo {year} {2009})}\BibitemShut {NoStop}%
\bibitem [{\citenamefont {Yasin}\ \emph {et~al.}(2011)\citenamefont {Yasin},
  \citenamefont {Dumm}, \citenamefont {Salameh}, \citenamefont {Batail},
  \citenamefont {Mezi{\'e}re},\ and\ \citenamefont {Dressel}}]{Yasin11}%
  \BibitemOpen
  \bibfield  {author} {\bibinfo {author} {\bibfnamefont {S.}~\bibnamefont
  {Yasin}}, \bibinfo {author} {\bibfnamefont {M.}~\bibnamefont {Dumm}},
  \bibinfo {author} {\bibfnamefont {B.}~\bibnamefont {Salameh}}, \bibinfo
  {author} {\bibfnamefont {P.}~\bibnamefont {Batail}}, \bibinfo {author}
  {\bibfnamefont {C.}~\bibnamefont {Mezi{\'e}re}},\ and\ \bibinfo {author}
  {\bibfnamefont {M.}~\bibnamefont {Dressel}},\ }\bibfield  {title} {\bibinfo
  {title} {{Transport studies at the Mott transition of the two-dimensional
  organic metal $\kappa$-{\rm
  (BEDT-TTF)$_2$Cu[N(CN)$_2$]Br$_{x}$Cl$_{1-x}$}}},\ }\href
  {https://doi.org/10.1140/epjb/e2010-10743-2} {\bibfield  {journal} {\bibinfo
  {journal} {Eur. Phys. J. B}\ }\textbf {\bibinfo {volume} {79}},\ \bibinfo
  {pages} {383} (\bibinfo {year} {2011})}\BibitemShut {NoStop}%
\bibitem [{\citenamefont {Geiser}\ \emph {et~al.}(1991)\citenamefont {Geiser},
  \citenamefont {Schultz}, \citenamefont {Wang}, \citenamefont {Watkins},
  \citenamefont {Stupka}, \citenamefont {Williams}, \citenamefont {Schirber},
  \citenamefont {Overmyer}, \citenamefont {Jung}, \citenamefont {Novoa},\ and\
  \citenamefont {Whangbo}}]{91GEISER-PhysC}%
  \BibitemOpen
  \bibfield  {author} {\bibinfo {author} {\bibfnamefont {U.}~\bibnamefont
  {Geiser}}, \bibinfo {author} {\bibfnamefont {A.~J.}\ \bibnamefont {Schultz}},
  \bibinfo {author} {\bibfnamefont {H.~H.}\ \bibnamefont {Wang}}, \bibinfo
  {author} {\bibfnamefont {D.~M.}\ \bibnamefont {Watkins}}, \bibinfo {author}
  {\bibfnamefont {D.~L.}\ \bibnamefont {Stupka}}, \bibinfo {author}
  {\bibfnamefont {J.~M.}\ \bibnamefont {Williams}}, \bibinfo {author}
  {\bibfnamefont {J.}~\bibnamefont {Schirber}}, \bibinfo {author}
  {\bibfnamefont {D.}~\bibnamefont {Overmyer}}, \bibinfo {author}
  {\bibfnamefont {D.}~\bibnamefont {Jung}}, \bibinfo {author} {\bibfnamefont
  {J.}~\bibnamefont {Novoa}},\ and\ \bibinfo {author} {\bibfnamefont {M.-H.}\
  \bibnamefont {Whangbo}},\ }\bibfield  {title} {\bibinfo {title} {{Strain
  index, lattice softness and superconductivity of organic donor-molecule
  salts: Crystal and electronic structures of three isostructural salts
  $\kappa$-(BEDT- TTF)$_2$Cu[N(CN)$_2$]$X$ ($X$=Cl, Br, I)}},\ }\href
  {https://doi.org/https://doi.org/10.1016/0921-4534(91)91586-S} {\bibfield
  {journal} {\bibinfo  {journal} {Physica C}\ }\textbf {\bibinfo {volume}
  {174}},\ \bibinfo {pages} {475 } (\bibinfo {year} {1991})}\BibitemShut
  {NoStop}%
\bibitem [{\citenamefont {Tanatar}\ \emph {et~al.}(2000)\citenamefont
  {Tanatar}, \citenamefont {Kagoshima}, \citenamefont {Ishiguro}, \citenamefont
  {Ito}, \citenamefont {Yefanov}, \citenamefont {Bondarenko}, \citenamefont
  {Kushch},\ and\ \citenamefont {Yagubskii}}]{00Tanatar-PhysRevB}%
  \BibitemOpen
  \bibfield  {author} {\bibinfo {author} {\bibfnamefont {M.~A.}\ \bibnamefont
  {Tanatar}}, \bibinfo {author} {\bibfnamefont {S.}~\bibnamefont {Kagoshima}},
  \bibinfo {author} {\bibfnamefont {T.}~\bibnamefont {Ishiguro}}, \bibinfo
  {author} {\bibfnamefont {H.}~\bibnamefont {Ito}}, \bibinfo {author}
  {\bibfnamefont {V.~S.}\ \bibnamefont {Yefanov}}, \bibinfo {author}
  {\bibfnamefont {V.~A.}\ \bibnamefont {Bondarenko}}, \bibinfo {author}
  {\bibfnamefont {N.~D.}\ \bibnamefont {Kushch}},\ and\ \bibinfo {author}
  {\bibfnamefont {E.~B.}\ \bibnamefont {Yagubskii}},\ }\bibfield  {title}
  {\bibinfo {title} {{Electronic transport properties and structural
  transformations of $\kappa$-(BEDT-TTF)$_{2}$Cu[N(CN)$_{2}$]I}},\ }\href
  {https://doi.org/10.1103/PhysRevB.62.15561} {\bibfield  {journal} {\bibinfo
  {journal} {Phys. Rev. B}\ }\textbf {\bibinfo {volume} {62}},\ \bibinfo
  {pages} {15561} (\bibinfo {year} {2000})}\BibitemShut {NoStop}%
\bibitem [{\citenamefont {Tanatar}\ \emph {et~al.}(2002)\citenamefont
  {Tanatar}, \citenamefont {Ishiguro}, \citenamefont {Kagoshima}, \citenamefont
  {Kushch},\ and\ \citenamefont {Yagubskii}}]{Tanatar02}%
  \BibitemOpen
  \bibfield  {author} {\bibinfo {author} {\bibfnamefont {M.~A.}\ \bibnamefont
  {Tanatar}}, \bibinfo {author} {\bibfnamefont {T.}~\bibnamefont {Ishiguro}},
  \bibinfo {author} {\bibfnamefont {S.}~\bibnamefont {Kagoshima}}, \bibinfo
  {author} {\bibfnamefont {N.~D.}\ \bibnamefont {Kushch}},\ and\ \bibinfo
  {author} {\bibfnamefont {E.~B.}\ \bibnamefont {Yagubskii}},\ }\bibfield
  {title} {\bibinfo {title} {{Pressure-temperature phase diagram of the organic
  superconductor $\kappa$-(BEDT-TTF)$_2$Cu[N(CN)$_2$]I }},\ }\href@noop {}
  {\bibfield  {journal} {\bibinfo  {journal} {Phys. Rev. B}\ }\textbf {\bibinfo
  {volume} {65}},\ \bibinfo {eid} {064516} (\bibinfo {year}
  {2002})}\BibitemShut {NoStop}%
\bibitem [{\citenamefont {Kanoda}(1997)}]{Kanoda97a}%
  \BibitemOpen
  \bibfield  {author} {\bibinfo {author} {\bibfnamefont {K.}~\bibnamefont
  {Kanoda}},\ }\bibfield  {title} {\bibinfo {title} {Recent progress in {{NMR}}
  studies on organic conductors},\ }\href
  {https://doi.org/10.1023/A:1012696314318} {\bibfield  {journal} {\bibinfo
  {journal} {Hyperfine Interact.}\ }\textbf {\bibinfo {volume} {104}},\
  \bibinfo {pages} {235} (\bibinfo {year} {1997})}\BibitemShut {NoStop}%
\bibitem [{\citenamefont {Mori}\ \emph {et~al.}(1999)\citenamefont {Mori},
  \citenamefont {Mori},\ and\ \citenamefont {Tanaka}}]{Mori99b}%
  \BibitemOpen
  \bibfield  {author} {\bibinfo {author} {\bibfnamefont {T.}~\bibnamefont
  {Mori}}, \bibinfo {author} {\bibfnamefont {H.}~\bibnamefont {Mori}},\ and\
  \bibinfo {author} {\bibfnamefont {S.}~\bibnamefont {Tanaka}},\ }\bibfield
  {title} {\bibinfo {title} {{Structural Genealogy of {\rm BEDT-TTF}-Based
  Organic Conductors II. Inclined Molecules: $\theta$, $\alpha$ and $\kappa$
  Phases}},\ }\href@noop {} {\bibfield  {journal} {\bibinfo  {journal} {Bull.
  Chem. Soc. Jpn.}\ }\textbf {\bibinfo {volume} {72}},\ \bibinfo {pages} {179}
  (\bibinfo {year} {1999})}\BibitemShut {NoStop}%
\bibitem [{\citenamefont {Furukawa}\ \emph {et~al.}(2015)\citenamefont
  {Furukawa}, \citenamefont {Miyagawa}, \citenamefont {Itou}, \citenamefont
  {Ito}, \citenamefont {Taniguchi}, \citenamefont {Saito}, \citenamefont
  {Iguchi}, \citenamefont {Sasaki},\ and\ \citenamefont
  {Kanoda}}]{15Furukawa-PhysRevLett}%
  \BibitemOpen
  \bibfield  {author} {\bibinfo {author} {\bibfnamefont {T.}~\bibnamefont
  {Furukawa}}, \bibinfo {author} {\bibfnamefont {K.}~\bibnamefont {Miyagawa}},
  \bibinfo {author} {\bibfnamefont {T.}~\bibnamefont {Itou}}, \bibinfo {author}
  {\bibfnamefont {M.}~\bibnamefont {Ito}}, \bibinfo {author} {\bibfnamefont
  {H.}~\bibnamefont {Taniguchi}}, \bibinfo {author} {\bibfnamefont
  {M.}~\bibnamefont {Saito}}, \bibinfo {author} {\bibfnamefont
  {S.}~\bibnamefont {Iguchi}}, \bibinfo {author} {\bibfnamefont
  {T.}~\bibnamefont {Sasaki}},\ and\ \bibinfo {author} {\bibfnamefont
  {K.}~\bibnamefont {Kanoda}},\ }\bibfield  {title} {\bibinfo {title} {Quantum
  spin liquid emerging from antiferromagnetic order by introducing disorder},\
  }\href {https://doi.org/10.1103/PhysRevLett.115.077001} {\bibfield  {journal}
  {\bibinfo  {journal} {Phys. Rev. Lett.}\ }\textbf {\bibinfo {volume} {115}},\
  \bibinfo {pages} {077001} (\bibinfo {year} {2015})}\BibitemShut {NoStop}%
\bibitem [{\citenamefont {Miksch}\ \emph {et~al.}(2020)\citenamefont {Miksch},
  \citenamefont {{Javaheri~Rahim}}, \citenamefont {Bardin}, \citenamefont
  {Kanoda}, \citenamefont {Schlueter}, \citenamefont {H\"{u}bner},
  \citenamefont {Scheffler},\ and\ \citenamefont {Dressel}}]{20Miksch}%
  \BibitemOpen
  \bibfield  {author} {\bibinfo {author} {\bibfnamefont {B.}~\bibnamefont
  {Miksch}}, \bibinfo {author} {\bibfnamefont {M.}~\bibnamefont
  {{Javaheri~Rahim}}}, \bibinfo {author} {\bibfnamefont {A.~A.}\ \bibnamefont
  {Bardin}}, \bibinfo {author} {\bibfnamefont {K.}~\bibnamefont {Kanoda}},
  \bibinfo {author} {\bibfnamefont {J.~A.}\ \bibnamefont {Schlueter}}, \bibinfo
  {author} {\bibfnamefont {R.}~\bibnamefont {H\"{u}bner}}, \bibinfo {author}
  {\bibfnamefont {M.}~\bibnamefont {Scheffler}},\ and\ \bibinfo {author}
  {\bibfnamefont {M.}~\bibnamefont {Dressel}},\ }\bibfield  {title} {\bibinfo
  {title} {{Gapped magnetic ground state in quantum-spin-liquid candidate
  $\kappa$-(BEDT-TTF)$_2$Cu$_2$(CN)$_3$}},\ }\href@noop {} {\bibfield
  {journal} {\bibinfo  {journal} {arXiv:2010.16155}\ } (\bibinfo {year}
  {2020})}\BibitemShut {NoStop}%
\bibitem [{\citenamefont {Antal}\ \emph {et~al.}(2009)\citenamefont {Antal},
  \citenamefont {Feh\'er}, \citenamefont {J\'anossy}, \citenamefont
  {T\'atrai-Szekeres},\ and\ \citenamefont {F\"ul\"op}}]{09Antal-PhysRevLett}%
  \BibitemOpen
  \bibfield  {author} {\bibinfo {author} {\bibfnamefont {{\'A}.}~\bibnamefont
  {Antal}}, \bibinfo {author} {\bibfnamefont {T.}~\bibnamefont {Feh\'er}},
  \bibinfo {author} {\bibfnamefont {A.}~\bibnamefont {J\'anossy}}, \bibinfo
  {author} {\bibfnamefont {E.}~\bibnamefont {T\'atrai-Szekeres}},\ and\
  \bibinfo {author} {\bibfnamefont {F.}~\bibnamefont {F\"ul\"op}},\ }\bibfield
  {title} {\bibinfo {title} {{Spin Diffusion and Magnetic Eigenoscillations
  Confined to Single Molecular Layers in the Organic Conductors
  $\kappa$-(BEDT-TTF)$_{2}$Cu[N(CN)$_{2}$]$X$ ($X$ = Cl, Br)}},\ }\href
  {https://doi.org/10.1103/PhysRevLett.102.086404} {\bibfield  {journal}
  {\bibinfo  {journal} {Phys. Rev. Lett.}\ }\textbf {\bibinfo {volume} {102}},\
  \bibinfo {pages} {086404} (\bibinfo {year} {2009})}\BibitemShut {NoStop}%
\bibitem [{\citenamefont {Antal}\ \emph {et~al.}(2010)\citenamefont {Antal},
  \citenamefont {Feh{\'e}r}, \citenamefont {N{\'a}fr{\'a}di}, \citenamefont
  {Ga{\'a}l}, \citenamefont {Forr{\'o}},\ and\ \citenamefont
  {J{\'a}nossy}}]{10Antal-PhysicaB}%
  \BibitemOpen
  \bibfield  {author} {\bibinfo {author} {\bibfnamefont {{\'A}.}~\bibnamefont
  {Antal}}, \bibinfo {author} {\bibfnamefont {T.}~\bibnamefont {Feh{\'e}r}},
  \bibinfo {author} {\bibfnamefont {B.}~\bibnamefont {N{\'a}fr{\'a}di}},
  \bibinfo {author} {\bibfnamefont {R.}~\bibnamefont {Ga{\'a}l}}, \bibinfo
  {author} {\bibfnamefont {L.}~\bibnamefont {Forr{\'o}}},\ and\ \bibinfo
  {author} {\bibfnamefont {A.}~\bibnamefont {J{\'a}nossy}},\ }\bibfield
  {title} {\bibinfo {title} {{Measurement of interlayer spin diffusion in the
  organic conductor $\kappa$-(BEDT-TTF)$_{2}$Cu[N(CN)$_{2}$]$X$, $X$ = Cl,
  Br}},\ }\href {https://doi.org/https://doi.org/10.1016/j.physb.2009.11.024}
  {\bibfield  {journal} {\bibinfo  {journal} {Physica B}\ }\textbf {\bibinfo
  {volume} {405}},\ \bibinfo {pages} {S168 } (\bibinfo {year}
  {2010})}\BibitemShut {NoStop}%
\bibitem [{\citenamefont {Antal}\ \emph {et~al.}(2011)\citenamefont {Antal},
  \citenamefont {Feh\'er}, \citenamefont {T\'atrai-Szekeres}, \citenamefont
  {F\"ul\"op}, \citenamefont {N\'afr\'adi}, \citenamefont {Forr\'o},\ and\
  \citenamefont {J\'anossy}}]{11Antal-PhysRevB}%
  \BibitemOpen
  \bibfield  {author} {\bibinfo {author} {\bibfnamefont {{\'A}.}~\bibnamefont
  {Antal}}, \bibinfo {author} {\bibfnamefont {T.}~\bibnamefont {Feh\'er}},
  \bibinfo {author} {\bibfnamefont {E.}~\bibnamefont {T\'atrai-Szekeres}},
  \bibinfo {author} {\bibfnamefont {F.}~\bibnamefont {F\"ul\"op}}, \bibinfo
  {author} {\bibfnamefont {B.}~\bibnamefont {N\'afr\'adi}}, \bibinfo {author}
  {\bibfnamefont {L.}~\bibnamefont {Forr\'o}},\ and\ \bibinfo {author}
  {\bibfnamefont {A.}~\bibnamefont {J\'anossy}},\ }\bibfield  {title} {\bibinfo
  {title} {{Pressure and temperature dependence of interlayer spin diffusion
  and electrical conductivity in the layered organic conductors
  $\kappa$-(BEDT-TTF)$_{2}$Cu[N(CN)$_{2}$]$X$ ($X$ = Cl, Br)}},\ }\href
  {https://doi.org/10.1103/PhysRevB.84.075124} {\bibfield  {journal} {\bibinfo
  {journal} {Phys. Rev. B}\ }\textbf {\bibinfo {volume} {84}},\ \bibinfo
  {pages} {075124} (\bibinfo {year} {2011})}\BibitemShut {NoStop}%
\bibitem [{\citenamefont {Kobayashi}\ \emph {et~al.}(2019)\citenamefont
  {Kobayashi}, \citenamefont {Suzuta}, \citenamefont {Tsuji}, \citenamefont
  {Ihara},\ and\ \citenamefont {Kawamoto}}]{19Kobayashi-PhysRevB}%
  \BibitemOpen
  \bibfield  {author} {\bibinfo {author} {\bibfnamefont {T.}~\bibnamefont
  {Kobayashi}}, \bibinfo {author} {\bibfnamefont {A.}~\bibnamefont {Suzuta}},
  \bibinfo {author} {\bibfnamefont {K.}~\bibnamefont {Tsuji}}, \bibinfo
  {author} {\bibfnamefont {Y.}~\bibnamefont {Ihara}},\ and\ \bibinfo {author}
  {\bibfnamefont {A.}~\bibnamefont {Kawamoto}},\ }\bibfield  {title} {\bibinfo
  {title} {{Inhomogeneous electronic state of organic conductor
  $\ensuremath{\kappa}\text{\ensuremath{-}}{(\text{BEDT-TTF})}_{2}{\mathrm{Cu}[\mathrm{N}(\mathrm{CN})}_{2}]\mathrm{I}$
  studied by $^{13}\mathrm{C}$ NMR spectroscopy}},\ }\href
  {https://doi.org/10.1103/PhysRevB.100.195115} {\bibfield  {journal} {\bibinfo
   {journal} {Phys. Rev. B}\ }\textbf {\bibinfo {volume} {100}},\ \bibinfo
  {pages} {195115} (\bibinfo {year} {2019})}\BibitemShut {NoStop}%
\bibitem [{\citenamefont {Dressel}\ and\ \citenamefont
  {Drichko}(2004)}]{Dressel04}%
  \BibitemOpen
  \bibfield  {author} {\bibinfo {author} {\bibfnamefont {M.}~\bibnamefont
  {Dressel}}\ and\ \bibinfo {author} {\bibfnamefont {N.}~\bibnamefont
  {Drichko}},\ }\bibfield  {title} {\bibinfo {title} {Optical properties of
  two-dimensional organic conductors: Signatures of charge ordering and
  correlation effects},\ }\href@noop {} {\bibfield  {journal} {\bibinfo
  {journal} {Chem. Rev.}\ }\textbf {\bibinfo {volume} {104}},\ \bibinfo {pages}
  {5689} (\bibinfo {year} {2004})}\BibitemShut {NoStop}%
\bibitem [{\citenamefont {Kataev}\ \emph
  {et~al.}(1992{\natexlab{a}})\citenamefont {Kataev}, \citenamefont {Winkel},
  \citenamefont {Knauf}, \citenamefont {Gruetz}, \citenamefont {Khomskii},
  \citenamefont {Wohlleben}, \citenamefont {Crump}, \citenamefont {Hahn},\ and\
  \citenamefont {Tebbe}}]{92KATAEV-PhysicaB}%
  \BibitemOpen
  \bibfield  {author} {\bibinfo {author} {\bibfnamefont {V.}~\bibnamefont
  {Kataev}}, \bibinfo {author} {\bibfnamefont {G.}~\bibnamefont {Winkel}},
  \bibinfo {author} {\bibfnamefont {N.}~\bibnamefont {Knauf}}, \bibinfo
  {author} {\bibfnamefont {A.}~\bibnamefont {Gruetz}}, \bibinfo {author}
  {\bibfnamefont {D.}~\bibnamefont {Khomskii}}, \bibinfo {author}
  {\bibfnamefont {D.}~\bibnamefont {Wohlleben}}, \bibinfo {author}
  {\bibfnamefont {W.}~\bibnamefont {Crump}}, \bibinfo {author} {\bibfnamefont
  {J.}~\bibnamefont {Hahn}},\ and\ \bibinfo {author} {\bibfnamefont
  {K.}~\bibnamefont {Tebbe}},\ }\bibfield  {title} {\bibinfo {title} {{ESR
  study of the electronic properties of the new organic conductors
  $\kappa$-(BEDT-TTF)$_2$Cu[N(CN)$_2$]$X$, $X$ = Br; I}},\ }\href
  {https://doi.org/https://doi.org/10.1016/0921-4526(92)90616-Z} {\bibfield
  {journal} {\bibinfo  {journal} {Physica B}\ }\textbf {\bibinfo {volume}
  {179}},\ \bibinfo {pages} {24 } (\bibinfo {year}
  {1992}{\natexlab{a}})}\BibitemShut {NoStop}%
\bibitem [{\citenamefont {Kataev}\ \emph
  {et~al.}(1992{\natexlab{b}})\citenamefont {Kataev}, \citenamefont {Winkel},
  \citenamefont {Khomskii}, \citenamefont {Wohlleben}, \citenamefont {Crump},
  \citenamefont {Tebbe},\ and\ \citenamefont {Hahn}}]{92KATAEV-SolStatCom}%
  \BibitemOpen
  \bibfield  {author} {\bibinfo {author} {\bibfnamefont {V.}~\bibnamefont
  {Kataev}}, \bibinfo {author} {\bibfnamefont {G.}~\bibnamefont {Winkel}},
  \bibinfo {author} {\bibfnamefont {D.}~\bibnamefont {Khomskii}}, \bibinfo
  {author} {\bibfnamefont {D.}~\bibnamefont {Wohlleben}}, \bibinfo {author}
  {\bibfnamefont {W.}~\bibnamefont {Crump}}, \bibinfo {author} {\bibfnamefont
  {K.}~\bibnamefont {Tebbe}},\ and\ \bibinfo {author} {\bibfnamefont
  {J.}~\bibnamefont {Hahn}},\ }\bibfield  {title} {\bibinfo {title} {{ESR of
  single crystals of $\kappa$-(BEDT-TTF)$_2$Cu[N(CN)$_2$]$X$ ($X$ = Br and
  I)}},\ }\href {https://doi.org/https://doi.org/10.1016/0038-1098(92)90084-M}
  {\bibfield  {journal} {\bibinfo  {journal} {Solid State Commun.}\ }\textbf
  {\bibinfo {volume} {83}},\ \bibinfo {pages} {435 } (\bibinfo {year}
  {1992}{\natexlab{b}})}\BibitemShut {NoStop}%
\bibitem [{Note1()}]{Note1}%
  \BibitemOpen
  \bibinfo {note} {This follows the elegant phenomenological description of
  lineshapes by Kubo \cite {Kubo69}; the two-state jump model it is also
  applied to model the splitting of vibrational lines in charge-order compounds
  with significant charge fluctuations \cite {Girlando12}}\BibitemShut
  {NoStop}%
\bibitem [{\citenamefont {Antal}(2011)}]{Thesis-Anges-Dissertation}%
  \BibitemOpen
  \bibfield  {author} {\bibinfo {author} {\bibfnamefont {{\'A}.}~\bibnamefont
  {Antal}},\ }\emph {\bibinfo {title} {Two dimensional spin transport and
  magnetism in layered organic crystals}},\ \href@noop {} {\bibinfo {type}
  {Dissertation}},\ \bibinfo  {school} {Budapest University of Technology and
  Economics}, \bibinfo {address} {Budapest, BME} (\bibinfo {year}
  {2011})\BibitemShut {NoStop}%
\bibitem [{\citenamefont {Kumar}\ and\ \citenamefont
  {Jayannavar}(1992)}]{92Kumar-PhysRevB}%
  \BibitemOpen
  \bibfield  {author} {\bibinfo {author} {\bibfnamefont {N.}~\bibnamefont
  {Kumar}}\ and\ \bibinfo {author} {\bibfnamefont {A.~M.}\ \bibnamefont
  {Jayannavar}},\ }\bibfield  {title} {\bibinfo {title} {Temperature dependence
  of the c-axis resistivity of high-${\mathit{t}}_{\mathit{c}}$ layered
  oxides},\ }\href {https://doi.org/10.1103/PhysRevB.45.5001} {\bibfield
  {journal} {\bibinfo  {journal} {Phys. Rev. B}\ }\textbf {\bibinfo {volume}
  {45}},\ \bibinfo {pages} {5001} (\bibinfo {year} {1992})}\BibitemShut
  {NoStop}%
\bibitem [{\citenamefont {Zverev}\ \emph {et~al.}(2006)\citenamefont {Zverev},
  \citenamefont {Manakov}, \citenamefont {Khasanov}, \citenamefont {Shibaeva},
  \citenamefont {Kushch}, \citenamefont {Kazakova}, \citenamefont {Buravov},
  \citenamefont {Yagubskii},\ and\ \citenamefont {Canadell}}]{06Zverev}%
  \BibitemOpen
  \bibfield  {author} {\bibinfo {author} {\bibfnamefont {V.~N.}\ \bibnamefont
  {Zverev}}, \bibinfo {author} {\bibfnamefont {A.~I.}\ \bibnamefont {Manakov}},
  \bibinfo {author} {\bibfnamefont {S.~S.}\ \bibnamefont {Khasanov}}, \bibinfo
  {author} {\bibfnamefont {R.~P.}\ \bibnamefont {Shibaeva}}, \bibinfo {author}
  {\bibfnamefont {N.~D.}\ \bibnamefont {Kushch}}, \bibinfo {author}
  {\bibfnamefont {A.~V.}\ \bibnamefont {Kazakova}}, \bibinfo {author}
  {\bibfnamefont {L.~I.}\ \bibnamefont {Buravov}}, \bibinfo {author}
  {\bibfnamefont {E.~B.}\ \bibnamefont {Yagubskii}},\ and\ \bibinfo {author}
  {\bibfnamefont {E.}~\bibnamefont {Canadell}},\ }\bibfield  {title} {\bibinfo
  {title} {Transport properties and structural features of the ambient-pressure
  superconductor $\kappa^{\prime}$-{{\rm (BEDT-TTF)$_{2}$Cu[N(CN)$_{2}$]Cl}}},\
  }\href {https://doi.org/10.1103/PhysRevB.74.104504} {\bibfield  {journal}
  {\bibinfo  {journal} {Phys. Rev. B}\ }\textbf {\bibinfo {volume} {74}},\
  \bibinfo {pages} {104504} (\bibinfo {year} {2006})}\BibitemShut {NoStop}%
\bibitem [{\citenamefont {{L.I. Buravov}}\ \emph {et~al.}(1992)\citenamefont
  {{L.I. Buravov}}, \citenamefont {{N.D. Kushch}}, \citenamefont {{V.A.
  Merzhanov}}, \citenamefont {{M.V. Osherov}}, \citenamefont {{A.G.
  Khomenko}},\ and\ \citenamefont {{E.B. Yagubskii}}}]{92Buraviv}%
  \BibitemOpen
  \bibfield  {author} {\bibinfo {author} {\bibnamefont {{L.I. Buravov}}},
  \bibinfo {author} {\bibnamefont {{N.D. Kushch}}}, \bibinfo {author}
  {\bibnamefont {{V.A. Merzhanov}}}, \bibinfo {author} {\bibnamefont {{M.V.
  Osherov}}}, \bibinfo {author} {\bibnamefont {{A.G. Khomenko}}},\ and\
  \bibinfo {author} {\bibnamefont {{E.B. Yagubskii}}},\ }\bibfield  {title}
  {\bibinfo {title} {Anisotropic resistivity and thermopower of the organic
  superconductor {{(ET)$_2$Cu[N(CN)$_2$]Br}}},\ }\href
  {https://doi.org/10.1051/jp1:1992207} {\bibfield  {journal} {\bibinfo
  {journal} {J. Phys. I (France)}\ }\textbf {\bibinfo {volume} {2}},\ \bibinfo
  {pages} {1257} (\bibinfo {year} {1992})}\BibitemShut {NoStop}%
\bibitem [{\citenamefont {Kushch}\ \emph {et~al.}(2001)\citenamefont {Kushch},
  \citenamefont {Tanatar}, \citenamefont {Yagubskii},\ and\ \citenamefont
  {Ishiguro}}]{Kushch01}%
  \BibitemOpen
  \bibfield  {author} {\bibinfo {author} {\bibfnamefont {N.~D.}\ \bibnamefont
  {Kushch}}, \bibinfo {author} {\bibfnamefont {M.~A.}\ \bibnamefont {Tanatar}},
  \bibinfo {author} {\bibfnamefont {E.~B.}\ \bibnamefont {Yagubskii}},\ and\
  \bibinfo {author} {\bibfnamefont {T.}~\bibnamefont {Ishiguro}},\ }\bibfield
  {title} {\bibinfo {title} {Superconductivity of
  $\kappa$-{(BEDT-TTF)$_2$Cu[N(CN)$_2$]I} under pressure},\ }\href
  {https://doi.org/https://doi.org/10.1134/1.1381643} {\bibfield  {journal}
  {\bibinfo  {journal} {JETP Lett.}\ }\textbf {\bibinfo {volume} {73}},\
  \bibinfo {pages} {429 } (\bibinfo {year} {2001})}\BibitemShut {NoStop}%
\bibitem [{\citenamefont {Kubo}(1969)}]{Kubo69}%
  \BibitemOpen
  \bibfield  {author} {\bibinfo {author} {\bibfnamefont {R.}~\bibnamefont
  {Kubo}},\ }\bibfield  {title} {\bibinfo {title} {A stochastic theory of line
  shapes},\ }\href {https://doi.org/10.1002/9780470143605.ch6} {\bibfield
  {journal} {\bibinfo  {journal} {Adv. Chem Phys.}\ }\textbf {\bibinfo {volume}
  {15}},\ \bibinfo {pages} {101} (\bibinfo {year} {1969})}\BibitemShut
  {NoStop}%
\bibitem [{\citenamefont {Girlando}\ \emph {et~al.}(2012)\citenamefont
  {Girlando}, \citenamefont {Masino}, \citenamefont {Kaiser}, \citenamefont
  {Sun}, \citenamefont {Drichko}, \citenamefont {Dressel},\ and\ \citenamefont
  {Mori}}]{Girlando12}%
  \BibitemOpen
  \bibfield  {author} {\bibinfo {author} {\bibfnamefont {A.}~\bibnamefont
  {Girlando}}, \bibinfo {author} {\bibfnamefont {M.}~\bibnamefont {Masino}},
  \bibinfo {author} {\bibfnamefont {S.}~\bibnamefont {Kaiser}}, \bibinfo
  {author} {\bibfnamefont {Y.}~\bibnamefont {Sun}}, \bibinfo {author}
  {\bibfnamefont {N.}~\bibnamefont {Drichko}}, \bibinfo {author} {\bibfnamefont
  {M.}~\bibnamefont {Dressel}},\ and\ \bibinfo {author} {\bibfnamefont
  {H.}~\bibnamefont {Mori}},\ }\bibfield  {title} {\bibinfo {title}
  {Spectroscopic characterization of charge order fluctuations in {{BEDT-TTF}}
  metals and superconductors},\ }\href {https://doi.org/10.1002/pssb.201100722}
  {\bibfield  {journal} {\bibinfo  {journal} {phys. stat. sol. (b)}\ }\textbf
  {\bibinfo {volume} {249}},\ \bibinfo {pages} {953} (\bibinfo {year}
  {2012})}\BibitemShut {NoStop}%
\end{thebibliography}%

\end{document}